\documentclass[10pt]{article}
\usepackage{amsmath}
\usepackage{amssymb}
\usepackage{graphics}
\usepackage{epsf,graphicx,rotating,color}
\usepackage{bm,bbm}
\usepackage{amsfonts}

\begin{document}
\title{{\bf  About One-Dimensional  Conservative Systems with Position Depending Mass}}
%
\author{G.V. L\'opez$^2$\thanks{gulopez@udgserv.cencar.udg.mx}\ ~ and C. Mart\'{\i}nez-Prieto$^1$\thanks{CARLOSR@iteso.mx}\\
\\\\$^1${\it \small Instituto Tecnol\'ogico y Estudios Superiores de Occidente,}\\
{\it \small  Perif\'erico Sur 8585l, CP 45604, Tlaquepaque, Jalisco, M\'exico.}  \\\\$^2${\it \small Departamento de F\'{\i}sica,
Universidad de Guadalajara, }\\ 
{\it\small  Blvd. Marcelino Garc\'{\i}a Barrag\'an 1421, esq. Calzada Ol\'{\i}mpica,}
\\{\it\small 44420
Guadalajara, Jalisco, M\'exico.}}
\date{April, 2014}
\maketitle
\noindent
\begin{center}
{\bf\Large Abstract}
\end{center}
For a one-dimensional conservative systems with position depending mass, one deduces 
consistently a constant of motion, a Lagrangian, and a Hamiltonian for the non relativistic case. With these functions, one shows the trajectories
on the spaces $(x,v)$ and ($x,p)$ for a linear position depending mass. For the relativistic case, the Lagrangian and
Hamiltonian can not be given explicitly in general. However, we study the particular system with constant force and mass linear dependence on the position where
the Lagrangian can be found explicitly,  but the Hamiltonian remains implicit in the constant of motion.
 \\Ê\\
{\bf keywords:} Mass variable systems, conservative system, position depending mass.
\setlength{\twocolumn}{}
\newpage
\section{\bf Introduction}
\noindent
Position depending mass systems have been relevant since the foundation of the classical mechanics and modern physics [1-5] (see reference there in). Actually, the interest for these type of problems has grown in modern physics due to fabrication of ultra thin semiconductors [6,7], inhomogeneous crystals [8], quantum dots [9], quantum liquids [10], and neutrino mass oscillations [11,12]. We also need to mention that this topic is important due to its relation with the foundation of the classical mechanics [13], and  its not invariance under Galileo or PoincarŽ-Lorentz transformations [13,14]. Most of the approaches  dealing with position depending mass problems use an intuitive way to write down a Lagrangian or Hamiltonian for the system, and then  solve the corresponding equations [15,16]. In this paper, one obtains a constant of motion , a Lagrangian, and a Hamiltonian in a consistent way for conservative non relativistic systems and study the harmonic oscillator with position depending mass as an example. For relativistic systems we point out the difficulty to get the same functions.   
\section{\bf Dynamical Functions}
A non relativistic conservative system with position depending mass is described by Newton's equation
\begin{equation}
\frac{d}{dt}\bigl(m(x)\dot x\bigr)=F(x),
\end{equation}     
where $\dot x$ denotes the velocity of the body with position depending mass $m(x)$, and with a force $F(x)$ acting on it . This problem can be written as the dynamical system
\begin{equation}
\dot x=v,\quad\quad\quad \dot v=\frac{F(x)}{m(x)}-\frac{m_x}{m} v^2,
\end{equation}
where $m_x$ is the differentiation of the mass with respect the position. A constant of motion of this system is a function $K=K(x,v)$ [17,18] which satisfies the following first order partial differential equation
\begin{equation}\label{eK}
v\frac{\partial K}{\partial x}+\frac{1}{m(x)}\biggl[F(x)-m_xv^2\biggr]\frac{\partial K}{\partial v}=0.
\end{equation} 
This equation can be solved by the characteristics method. 
The equations for its characteristics curves are
\begin{equation}\label{ch1}
\frac{dx}{v}=\frac{m(x)~ dv}{F(x)-m_xv^2}=\frac{dK}{0}.
\end{equation} 
From the last term, one knows that the solution of (\ref{eK}) must be of the form 
\begin{equation}
K(x,v)=G\bigl(C(x,v)\bigr),
\end{equation}
where $C(x,v)$ is the characteristic curve obtained from the first two terms of (\ref{ch1}), and being $G$ and arbitrary function. This characteristic curve can be found arranging these two terms of the form
\begin{equation}
m(x)v\frac{dv}{dt}=F(x)-m_xv^2,
\end{equation}
and defining a new variable, $\xi=v^2$, to get the equation
\begin{equation}
m(x) \frac{d\xi}{dx}+2m_x\xi=2F(x)
\end{equation}
which can readily be integrated to obtain the characteristic curve
\begin{equation}
C(x,v)=\frac{m^2(x)}{2}v^2-\int m(x)F(x)dx.
\end{equation}
Choosing the initial conditions $x(0)=0$, $v(0)\not=0$, and $m(0)=m_0$, and selecting the functionality $G(C(x,v))=C(x,v)/m_0$, the constant of motion is
\begin{equation}
K(x,v)=\frac{m^2(x)}{2m_0}v^2+V_{eff}(x),
\end{equation}
where $V_{eff}$ is the effective potential due to the position depending mass,
\begin{equation}\label{veff}
V_{eff}(x)=-\frac{1}{m_0}\int m(x)F(x)~dx,
\end{equation}
and this constant of motion has the right expression when constant mass is considered. 
Using now the known expression [19,20,21] to get the Lagrangian from a constant of motion,
\begin{equation}\label{La}
L(x,v)=v\int^v\frac{K(x,\xi)}{\xi^2}~d\xi,
\end{equation}
The Lagrangian, generalized linear momentum and Hamiltonian are given by
\begin{equation}
L(x,v)=\frac{m^2(x)}{2m_0}v^2-V_{eff}(x),
\end{equation}
\begin{equation}
p(x,v)=\frac{m^2(x)}{m_0}v,
\end{equation}
and
\begin{equation}
H(x,p)=\frac{m_0}{2m^2(x)}p^2+V_{eff}(x).
\end{equation}
The above expression for the dynamical functions show that there are two main modifications from the usual expression 
when position depending mass is considered. Firstly,  an effective potential is created which depends on $m(x)$ (\ref{veff}). Secondly, the kinetic energy like term is not of the form $p^2/2m(x)$ but it is of the form $m_0p^2/2m^2(x)$. These two modification are really important to deal correctly with an specific mass position depending conservative problem.
\section{Harmonic Oscillator $m(x)$}
Consider the harmonic oscillator, $F(x)=-kx$, with a linear position depending mass, $m(x)=m_0+m_{_1}x$. Thus, the effective potential, constant of motion, and Hamiltonian are given by
\begin{equation}
V_{eff}(x)=\frac{k}{2}x^2+\frac{m_{_1}k}{3m_0}x^3,
\end{equation}
\begin{equation}\label{kv1}
K(x,v)=\frac{(m_0+m_{_1}x)^2}{2m_0}v^2+\frac{k}{2}x^2+\frac{m_{_1}k}{3m_0}x^3,
\end{equation}
and
\begin{equation}\label{hp1}
H(x,p)=\frac{m_0}{2(m_0+m_{_1}x)^2}p^2+\frac{k}{2}x^2+\frac{m_{_1}k}{3m_0}x^3.
\end{equation}
Figures 1a and 1b show the trajectories on the spaces ($x,v$) and ($x,p$)  generated by the constant of motion and the Hamiltonian above with $m_0=200$ and for $m_{_1}>0$ ($m_0=200$; $m_{_1}=0$ (1), $m_{_1}=5$ (2), $m_{_1}=10$ (3), $m_{_1}=20$ (4)).  Figures 2a and 2b show also the trajectories on those spaces with $m_0=1000$ and $m_{_1}<0$ ($m_{_1}=0$ (1), $m_{_1}=-1$ (2), $m_{_1}=-1.5$ (3), $m_{_1}=-2$ (4)).  From Figure 2a one notes a singular behavior  in the velocity which comes from  (\ref{kv1}) and does not appears in the Hamiltonian formulation (\ref{hp1}). 
\begin{figure}[t]
\includegraphics[scale=0.3,angle=0]{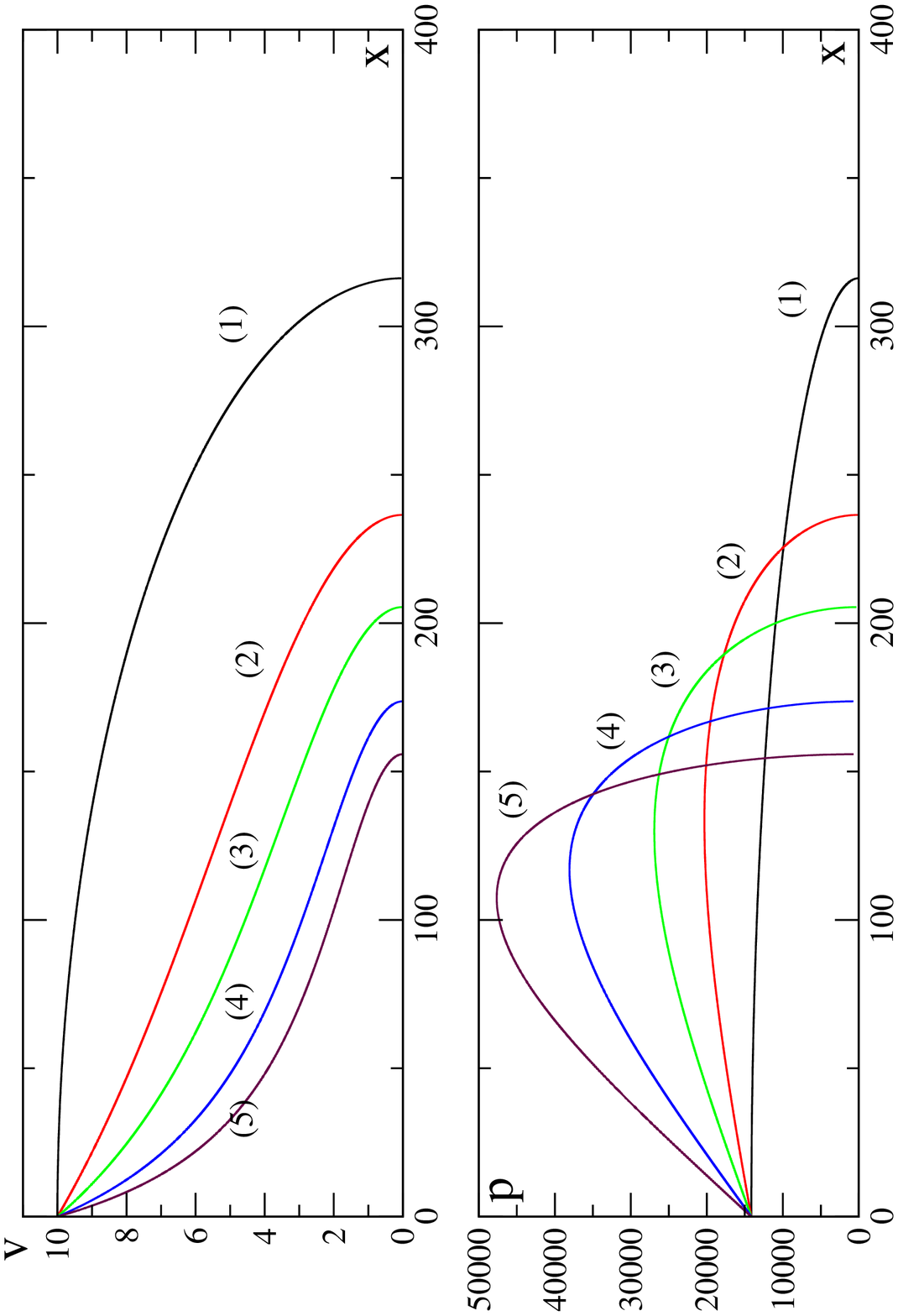}
    \caption{Trajectories for $m_{_1}>0$. }
    \label{fig:A}
\end{figure}
\\Ê\\
\begin{figure}[t]
\includegraphics[scale=0.3,angle=0]{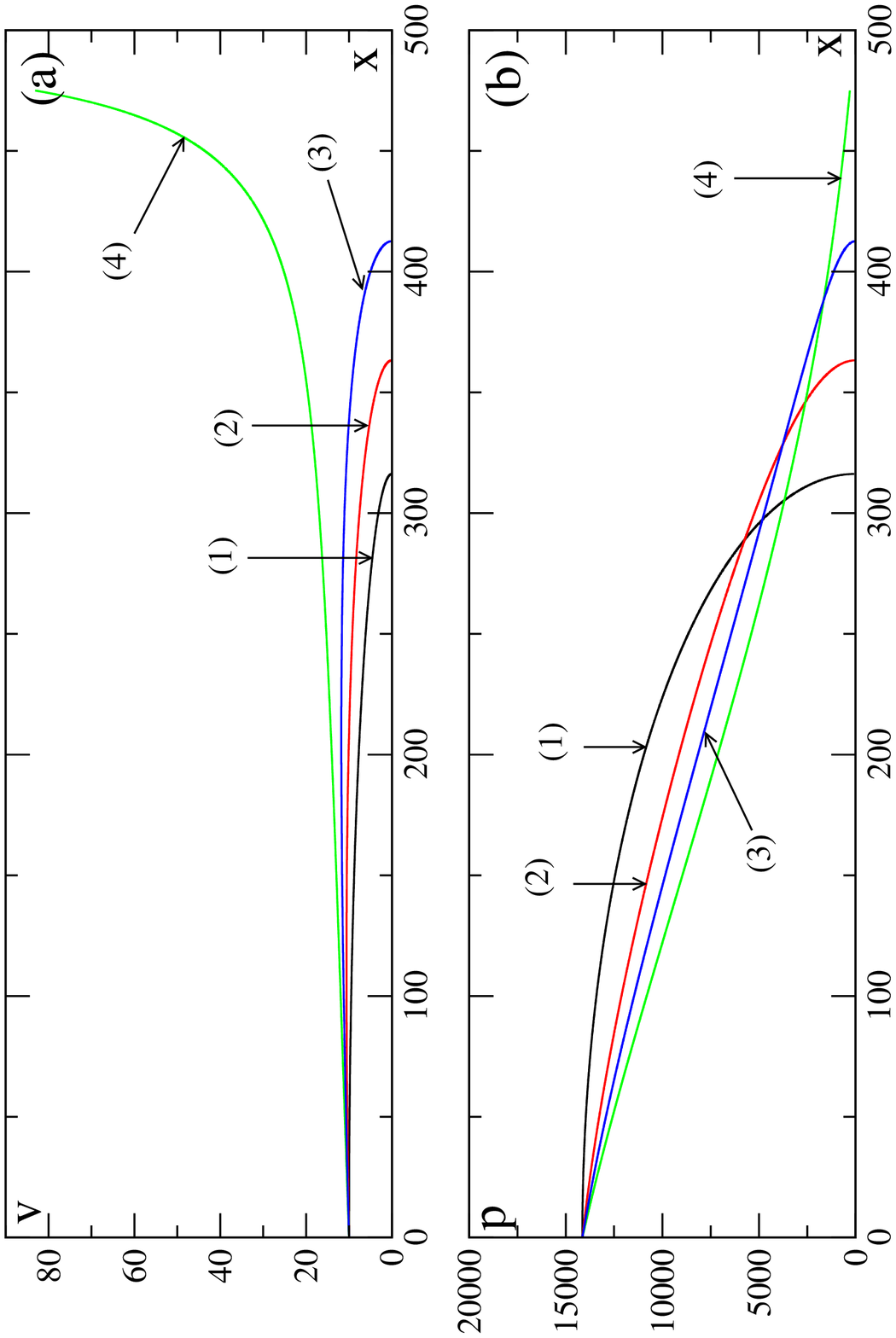}
    \caption{Trajectories for $m_{_1}<0$. }
    \label{fig:A}
\end{figure}
Note that due to relation (\ref{veff}), it is not possible to know the potential (effective) without the acknowledge of the position depending mass previously.  
\section{Relativistic case}
The relativistic motion [22] of a body with position depending mass is not invariant under Poincar\'e-Lorentz transformation [13], but it still can be described by the equation
\begin{equation}
\frac{d}{dt}(\gamma m(x) \dot x)=F(x),\quad\quad \gamma=(1-\frac{\dot x^2}{c^2})^{-1/2},
\end{equation}
where $c$ is the speed of light, and  it can be written as a Newton's equation with a velocity depending force of the form
\begin{equation}
m(x)\frac{d^2x}{dt^2}=\biggl[F(x)-\frac{ \dot x^2 m_x}{\sqrt{1-(\dot x/c)^2}}\biggr]\left(1-\frac{\dot x^2}{c^2}\right)^{3/2}.
\end{equation}
In turns, this equation defines the dynamical system
\begin{eqnarray}
& &\dot x=v, \\ \nonumber\\
& &\dot v=\frac{1}{m(x)}\biggl[F(x)-\gamma{ v^2 m_x}\biggr]\gamma^{-3}.
\end{eqnarray}
As before, a constant of motion of this autonomous system is a function $K=K(x,v)$ satisfying the equation
\begin{equation}
v\frac{\partial K}{\partial x}+\frac{1}{m(x)}\biggl[F(x)-\gamma{ v^2 m_x}\biggr]\gamma^{-3}\frac{\partial K}{\partial v}=0.
\end{equation}
The equations for the characteristics are
\begin{equation}
\frac{dx}{v}=\frac{m(x) dv}{\bigl[F(x)-\gamma{ v^2 m_x}\bigr]\left(1-\frac{v^2}{c^2}\right)^{3/2}}=\frac{dK}{0}.
\end{equation}
Thus, this constant of motion is an arbitrary function of the characteristic obtained from the first two term of this expression, $K=G(C(x,v))$, and from the first two terms one gets the following equation in terms of the  variable $\eta=v^2/c^2$,
\begin{equation}\label{qeta}
\frac{m(x) c^2}{2}\frac{d\eta}{dx}=\left[F(x)-\frac{m_xc^2\eta}{\sqrt{1-\eta}}\right]\left(1-\eta\right)^{3/2}.
\end{equation}
The constant of integration of this equation will represent the characteristic curve $C(x,\eta)$. Of course, in general the solution of this equation is not expressed in  close form. This means that the Lagrangian and the Hamiltonian of the system can not be found in general.
 However, there is a particular case where one can do something analytically, and this case consists of having a constant force  with mass linearly  dependence on the position.
\subsection{ Constant force with $m(x)=m_0+m_{_1}x$.}
In this particular case, one has that $m_x=m_{_1}$ and $F(x)=F=constant$, and the variables can be separated for the integration in (\ref{beta}), bringing about the characteristic curve
\begin{eqnarray}
& &C(x,\eta)=\int\frac{d\eta}{\displaystyle\left(F-\frac{m_{_1}\eta}{\sqrt{1-\eta}}\right)(1-\eta)^{3/2}}\nonumber\\ 
& &\quad -\frac{2}{c^2m_{_1}}\ln(m_0+m_{_1}x).
\end{eqnarray}
By selecting the constant of motion as
\begin{equation}
K(x,\eta)=\frac{Fm_0c^2}{2}C(x,\eta)+\frac{Fa}{m_{_1}}\ln m_0-m_0c^2,
\end{equation}
that is,
\begin{eqnarray}\label{c1}
& &K(x,\eta)=\frac{m_0c^2}{2}\int\frac{d\eta}{\left(1-\frac{m_{_1}\eta}{F\sqrt{1-\eta}}\right)(1-\eta)^{3/2}}\nonumber\\
& &\quad -\frac{m_0F}{m_{_1}}\ln(m_0+m_{_1}x)+\frac{Fm_0}{m_{_1}}\ln m_0-m_0c^2,\nonumber\\
\end{eqnarray}
one has the following limit
\begin{equation}
\lim_{m_{_1}\to 0}K(x,\eta)=\gamma m_0c^2-Fx-m_0c^2,
\end{equation}
which corresponds to the relativistic case of constant mass.
Now, considering the condition
\begin{equation}
\left|\frac{m_{_1}\eta}{F\sqrt{1-\eta}}\right|\le 1,
\end{equation}
one can write the constant of motion as [23]
\begin{eqnarray}
& &K(x,\eta)=\gamma m_0c^2-Fx-m_0c^2\nonumber\\
& &+m_{_1}\left\{\frac{Fx^2}{2m_0}+\frac{m_0c^2}{2}\int\frac{\eta d\eta}{(1-\eta)^2}\right\}\nonumber\\
& &+\frac{m_0c^2}{2}\sum_{k=3}^{\infty}\int\left(\frac{m_{_1}\eta}{F(\sqrt{1-\eta}}\right)^{k-1}\frac{d\eta}{(1-\eta)^{3/2}}\nonumber\\
& &+\sum_{k=3}^{\infty}\frac{(-1)^kFm_0}{m_{_1}k}\left(\frac{m_{_1}x}{m_0}\right)^k.
\end{eqnarray}
or
\begin{eqnarray}\label{rcm}
& &K(x,\eta)=\gamma m_0c^2-Fx-m_0c^2\nonumber\\
& &+m_{_1}\left\{\frac{Fx^2}{2m_0}+\frac{m_0c^2}{2}\left[\ln(1-\eta)+\frac{1}{1-\eta}\right]\right\}\nonumber\\
& &+\frac{m_0c^2}{2}\sum_{k=3}^{\infty}\int\left(\frac{m_{_1}\eta}{F(\sqrt{1-\eta}}\right)^{k-1}\frac{d\eta}{(1-\eta)^{3/2}}\nonumber\\
& &+\sum_{k=3}^{\infty}\frac{(-1)^kFm_0}{m_{_1}k}\left(\frac{m_{_1}x}{m_0}\right)^k,
\end{eqnarray}
where the summation represents terms of order two or higher in the parameter $m_{_1}$.  
The Lagrangian of the system (ref{La}) in terms of the variable $\eta$ is
\begin{equation}
L(x,\eta)=\frac{\sqrt{\eta}}{2}\int^{\eta}\frac{K(x,\rho) d\rho}{\rho\sqrt{\rho}}.
\end{equation}
So, using the above constant of motion in this expression, one gets
\begin{eqnarray}
& &L(x,\eta)=-m_0c^2\sqrt{1-\eta}+Fx+m_0c^2\nonumber\\
& &+m_{_1}\biggl\{-\frac{Fx^2}{2m_0}+
\frac{m_0c^2}{2}\biggl[\frac{3\sqrt{\eta}}{2}\ln\left|\frac{-1+\sqrt{\eta}}{1+\sqrt{\eta}}\right|\nonumber\\
& &\quad\quad-\ln(1-\eta)-1\biggr]\biggr\}\nonumber\\
& &+\frac{m_0c^2\sqrt{\eta}}{4}\sum_{k=3}^{\infty}\int^{\eta}\frac{d\eta'}{\eta'\sqrt{\eta'}}\int^{\eta'}\frac{\left(\frac{m_{_1}\rho}{\sqrt{1-\rho}}\right)^{k-1} d\rho}{(1-\rho)^{3/2}}\nonumber\\
& &-\sum_{k=3}^{\infty}\frac{(-1)^kFm_0}{m_{_1}k}\left(\frac{m_{_1}x}{m_0}\right)^k\nonumber\\
\end{eqnarray}
The generalized linear momentum in terms of the variable $\eta$,
\begin{equation}
p=\frac{2\sqrt{\eta}}{c}\left(\frac{\partial L}{\partial \eta}\right),
\end{equation}
is given by
\begin{eqnarray}\label{rp}
& & p=\gamma m_0c\sqrt{\eta}\nonumber\\
& &+\frac{m_{_1}m_0c^2}{2}\left\{ \frac{1}{2c}\ln\left|\frac{-1+\sqrt{\eta}}{1+\sqrt{\eta}}\right|+\frac{\eta}{c(1-\eta)}\right\}\nonumber\\
& &+\frac{m_0c\sqrt{\eta}}{2}\sum_{k=3}^{\infty}\frac{\partial}{\partial\eta}\left(\sqrt{\eta}\int^{\eta}\frac{d\eta'}{\eta'\sqrt{\eta'}}\int^{\eta'}\frac{\left(\frac{m_{_1}\rho}{\sqrt{1-\rho}}\right)^{k-1} d\rho}{(1-\rho)^{3/2}}\right)\nonumber\\
\end{eqnarray}
Now, as one can see from this expression, even a first order in $m_1$ it is not possible to obtain $\eta=\eta(x,p)$ in order to get the Hamiltonian of the system. Therefore, one can not have explicitly the Hamiltonian of the system but it remains implicit  through the constant of motion (\ref{rcm}).
\section{Conclusion}
We have shown that mass position depending problems for non relativistic conservative systems bring about a modification to the potential  and kinetic energy terms. The constant of motion and Hamiltonian of these system differs greatly since the generalized linear momentum depends on the position and the velocity of the body. This differences are shown with the  trajectories on the spaces $(x,v)$ and $(x,p$). For the relativistic conservative systems with mass position depending, the full integrability is not so simple in general, but we analyzed the particular case of constant force with mass linear dependence on position system. As we showed, even we can get the Lagrangian for the system, it is not possible to obtain the inverse relation $v=v(x,p)$, and therefore, the Hamiltonian of this system.     

\clearpage

\end{document}